\documentclass[12pt]{iopart}
\usepackage{mathbbold}
\usepackage{graphicx}

\begin{document}

\title[Strange Quark stars: Observations \& Speculations]
{Strange Quark stars: Observations \& Speculations}

\author{Renxin Xu}

\address{School of Physics and State Key Laboratory of Nuclear
Physics and Technology, Peking University, Beijing 100871, China}
\ead{r.x.xu@pku.edu.cn}
\begin{abstract}

Two kinds of difficulties have challenged the physics community for
many years: (1) knowing nature's building blocks (particle physics)
and (2) understanding interacting many-body systems (many-body
physics).
Both of them exist in the research of quark matter and compact
stars.
This paper addresses the possibility that quark clustering, rather
than a color super-conducting state, could occur in cold quark
matter at realistic baryon densities of compact stars, since a
weakly coupling treatment of the interaction between quarks might
not be reliable. Cold quark matter is conjectured to be in a solid
state if thermal kinematic energy is much lower than the interaction
energy of quark clusters.
Different manifestations of pulsar-like compact stars are discussed,
as well as modeled in a regime of solid quark stars.
\end{abstract}

\vspace{5mm}%
\noindent%
{\bf 1. Introduction: what's the nature of pulsar-like stars?}

There are strange and interesting stories about success and failure
in the road to understand Nature.
As an example, we will briefly review some significant moments in
the research of neutron stars and pulsars.
In 1932, soon after Chandrasekhar found a unique mass (the mass
limit of white dwarfs), Landau speculated a state of matter, the
density of which ``becomes so great that atomic nuclei come in close
contact, forming one {\em gigantic nucleus}''.
A star composed mostly of such matter is called a ``neutron'' star,
and Baade and Zwicky even suggested in 1934 that neutron stars (NSs)
could be born after supernovae.
NSs theoretically predicted were finally {\em discovered} when
Hewish and his collaborators detected radio pulsars in 1967.
More kinds of pulsar-like stars, such as X-ray pulsars and X-ray
bursts in binary systems, were also discovered later, and all of
them are suggested to be NSs.

However, the simple and beautiful idea proposed by Landau and others
had one flaw at least:
nucleons (neutrons and protons) are in fact {\em not} structureless
point-like particles although they were thought to be elementary
particles in 1930s, they (and other hadrons) were proposed to be
composed of {\em quarks} already in the 60's!
Naively that question becomes: can the quark degree of freedoms
appear when nucleons approach each other in a compact star with
supra-nuclear density? The answer turned to be a little bit more
affirmative after physicists recognized that the elementary strong
interaction between quarks is actually asymptotically free, i.e.,
the interaction becomes weak (and quarks are thus deconfined) when
density and/or temperature are extremely high.
Therefore, quark matter could possibly exist in compact stars,
either in the cores of neutron stars (mixed or hybrid
stars~\cite{ik69}) or in the whole stars (quark
stars~\cite{itoh70,hzs86,afo86}).
Now, after more than 40 years since the discovery of the first
pulsar CP 1919, the real nature of pulsars is still uncertain
because of difficulties in both theories and observations (e.g.
\cite{weber05}).

\vspace{2mm}%
\noindent%
{\bf 2. Cold quark matter: color superconducting {\em vs} quark
clustering?}

Quark matter (or quark-gluon plasma) not only is a state predicted
theoretically in QCD (quantum chromodynamics), the underlying theory
of the elementary strong interaction, but also would be the key to
sub-quarkian physics.
Hot quark matter could be reproduced in the experiments of
relativistic heavy ion collisions, and hadronization occurs soon as
the fireball cools. The final states of hadrons recorded in
detectors are used to infer the hot quark matter.
On the other hand, extremely high chemical potential is required to
create cold quark matter, and it can only exist in rare
astrophysical conditions, the compact stars.
Cold quark matter is relatively long-lived, and in principle we
could reliably know its real state by astronomical observations
since there is possible evidence that pulsar-like stars are actually
quark stars.

What kind of cold matter can we expect from QCD theory, in effective
models, or even based on phenomenology?
This is a question too hard to answer because of (i) the
non-perturbative effect of strong interaction between quarks at low
energy scales and (ii) the many-body problem due to vast assemblies
of interacting particles.
A color-superconductivity (CSC) state is currently focused on in
QCD-based models, as well as in phenomenological ones (e.g.
\cite{csc08}). However, an alternative suggestion that cold quark
matter could be in a solid state~\cite{xu03,Horvath05,Owen05} can
not be ruled out yet in both astrophysics and particle physics.
Recently, based on an approximation scheme of a large number of
colors, $N_c$, a quarkyonic state of matter~\cite{Quarkyonic07} was
also proposed to exist at ultra-hight density, where a baryonic
``skin'' with depth of a few 100 MeV may form near the Fermi surface
due to strong color interaction.

Let's discuss this in more details, beginning with a brief
introduction to the essence of QCD related to strong interaction.
In quantum field theory, the strength of the interaction is measured
by a coupling parameter $g$ whose dependence on the energy-scale,
$\mu$, is determined by the relation of
$ \beta(g)\equiv \partial g/\partial \ln\mu=\mu \partial g/\partial \mu.%
$
The beta-function, $\beta(g)$, represents the running of coupling.
In an Abelian gauge theory as QED, the beta function is positive.
However, in a non-Abelian gauge theory $\beta$ is negative, and
hence, the QCD coupling decreases at high energies, which goes
approximately as~\cite{running}
\begin{equation}
\alpha_{\rm s}(\mu)\equiv {g_{\rm s}^2\over 4\pi} \approx{1\over
\beta_0\ln(\mu^2/\Lambda^2)},%
\label{alpha}
\end{equation}
where $\beta_0=(11-2n_f/3)/(4\pi)$, $n_f$ is the number of quark
flavors, and the renormalization parameter $\Lambda=(200\sim 300)$
MeV. Certainly, the coupling is strong at low energies, and the
perturbative formulation is not applicable.

For cold dense quark matter, the order of the scale $\mu$ is
determined by the baryonic chemical potential,
$%
\mu\simeq (3\pi^2)^{1/3}\hbar c n_{\rm B}^{1/3}~%
$%
for $T\ll \mu$, where $T$ denotes the temperature, and $n_{\rm B}$
is the baryon number density. It is evident from Eq.(\ref{alpha})
that perturbative QCD is reliable in the limit of high density
($n_{\rm B}\rightarrow \infty$) because of asymptotic freedom.
The ground state of extremely dense quark matter could certainly be
that of an ideal Fermi gas. Nevertheless, it has been found that the
highly degenerate Fermi surface is unstable against the formation of
quark Cooper pairs, which condense near the Fermi surface due to the
existence of color-attractive channels between the quarks. A
BCS-like color superconductivity, similar to electric
superconductivity, has been formulated within perturbative QCD at
ultra-high baryon densities. It has been argued, based on QCD-like
effective models, that color superconductivity could also occur even
at the more realistic baryon densities of pulsar-like compact
stars~\cite{csc08}.

\begin{figure}
  \centering
    \includegraphics[width=5cm]{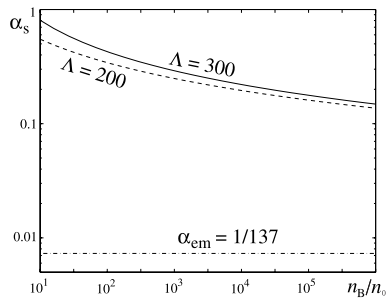}
    \caption{%
The running coupling, $\alpha_s$, in cold quark matter as a function
of baryon density, $n_{\rm B}$, for cut-off parameter $\Lambda=200$
(solid line) and 300 (dashed line), respectively. For comparison,
the usual electromagnetic coupling constat $\alpha_{\rm em}\simeq
1/137$ (dash-dotted line) is also drawn. A weakly coupling treatment
could be dangerous for cold quark matter at a realistic baryon
density ($\sim$ a few nuclear density, $n_0$).
\label{coupling}}
\end{figure}

Can the realistic stellar densities be high enough to justify the
use of perturbative QCD?
Let's see the numerical coupling strength from Eq.(\ref{alpha}),
shown in Fig.\ref{coupling}, with $n_f=3$ for strange quark matter.
It is observed that $\alpha_{\rm s}=(0.5\sim 0.8)$ when $n_{\rm
B}=10n_0$ ($n_0=0.16$ fm$^{-3}$), and $\alpha_{\rm s}\simeq 0.15$
even if $n_{\rm B}=10^6n_0$.
It is worth noting that the dimensionless electromagnetic coupling
constant (i.e., the fine-structure constant) is $1/137<0.01$, which
makes QED tractable.
That is to say, a weakly coupling strength comparable with that of
QED is possible in QCD only if the density is unbelievably and
unrealistically high ($n_{\rm B}>10^{123}n_0$!). At realistic
densities of a few nuclear density, $n_{\rm B}\geq n_0$, the color
coupling should be very strong rather weak, $\alpha_{\rm s}=(0.8\sim
1.5)$ for $n_{\rm B}=3n_0$, according to Eq.(\ref{alpha}).
This surely means that a weakly coupling treatment could be
dangerous for cold quark matter, i.e., the non-perturbative effect
in QCD should not be negligible if we try to know the real state of
compact stars.

\begin{figure}
  \centering
    \includegraphics[width=5cm]{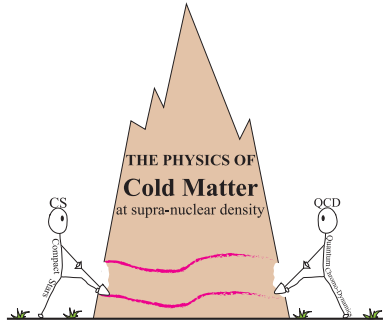}
    \caption{%
Astronomers interested in pulsar-like compact stars and physicists
expert at doing with QCD now face a same challenge to know the real
state of cold matter at supra-nuclear density. They have to exchange
information during their work when trying to ``dig a tunnel'' to
solve the problem.
\label{csqcd}}
\end{figure}

However, non-perturbative QCD is one of the daunting challenges
nowadays in understanding the fundamental strong interaction between
quarks.
Is there any other way for us to understand the physics of cold
dense matter at supra-nuclear density?
Compact stellar objects are the natural astrophysical laboratories
to probe the mystery of cold quark matter!
As addressed in \S1, pulsar-like compact stars could be quark stars,
and there is possible evidence for quark
stars~\cite{weber05,xu_huangshan,xu08_mpl}.
In a word, both physicists and astrophysicists are troubled with
cold dense matter, and they have to cooperate in order to completely
solve the problem, as illustrated in Fig.\ref{csqcd}. It is crucial
to develop a close communication between them.

What valuable information can astrophysics provide for us to know
the state of cold (quark) matter at supranuclear density?
Drifting sub-pulses of radio pulsars, non-atomic spectra of isolated
``neutron'' stars, and even successful supernova and $\gamma$-ray
bursts may hint to determine if pulsar-like compact stars could be
(bare) quark stars. Similarly, pulsar glitches, both normal and
slow, precessions, huge energy release of soft $\gamma$-ray
repeaters, and even the quantitative spectra of isolated ``neutron''
stars may suggest that such compact stars could globally be in a
solid state, i.e., they might be solid quark stars.
The details of these phenomena are presented and explained in the
next section.
The effects of the strong magnetic fields typically present in
neutron stars are ignored throughout this paper.  However, it is
known that in the case of color superconductivity these fields can
influence the ground state of the system~\cite{fdm05,fdm06}, produce
new phases~\cite{fd07a} or even trigger and/or cure chromomagnetic
instabilities~\cite{fd06,fd07b}. Hence, it remains an open question
whether the magnetic fields of the stars can also have relevant
effects on a solid state of quark matter (e.g. the
ferromagnetization~\cite{xu05}).

Can cold quark matter be in a solid state? This is possible because
of quark clustering~\cite{xu03} (normal solid) and the CSC
gap-parameter modulating~\cite{mrs07} (super solid).
Quark clusters may form in relatively low temperature
quark matter due to the strong interaction%
\footnote{%
As an example to show the effect of strong coupling between
particles on the state of particle system, we discuss a ``$p+e$''
system. If the interaction is turned off, the system is certainly a
Fermi gas with two degrees of freedom ($p$ and $e$), when the
temperature $T$ is so low that the de Broglie wavelength ($\propto
T^{-1/2}$) is $\geq$ the distance between particles. However, a
realistic ``$p+e$'' system at low temperature and density is
hydrogen (or even H$_2$) gas, i.e. particle clusters form due to
electric coupling there.
}%
, and the clusters could be in periodic lattices (normal solid) when
temperature becomes low enough.
Although it is hitherto impossible to know if quark clusters could
form in cold quark matter via calculation from first principles,
there could be a few points that favor clustering.
{\em Experimentally}, though quark matter is argued to be weakly
coupled at high energies and thus deconfined, it is worth noting
that, as revealed by the recent achievements in relativistic heavy
ion collision experiments, the interaction between quarks in a
fireball of quarks and gluons is still very strong (i.e. the
strongly coupled quark-gluon plasma, sQGP~\cite{Shuryak}).
The strong coupling between quarks may naturally render quarks
grouped in clusters, i.e., a condensation in position space rather
than in momentum space.
{\em Theoretically}, the baryon-like particles in quarkyonic
matter~\cite{Quarkyonic07} might be grouped further due to residual
color interaction if the baryon density is not extremely high, i.e.
$\mu >\sim \Lambda$, and quark clusters would form then.

In fact, strongly coupled particles are very complex systems, even
in the case of electromagnetic interaction~\cite{eQGP,eQGPa,eQGPb}.
It is recognized that in the BCS superfluid, fermions are condensed
in momentum space to form loosely bound Cooper pairs. On the other
hand, in the BEC (Bose-Einstein condensation), fermions are
condensed in position space to form tightly bound fermion pairs.
These are two limiting cases of same theory.
As for cold quark matter, although BCS-like CSC can occur at a
ultra-high density according to perturbative QCD, the interaction
becomes stronger and stronger as the density decreases, and the
quark pairs may get localized to create diquark boson (i.e., the
system crosses from BCS to BEC states).
Can diquarks condense further in position space to form
quark-clusters? This could be possible in cold quark matter at lower
density, an experimental analogy of which could be of the $^{87}$Rb
boson system~\cite{Rb87}.

\begin{figure}
  \centering
    \includegraphics[width=7.5cm]{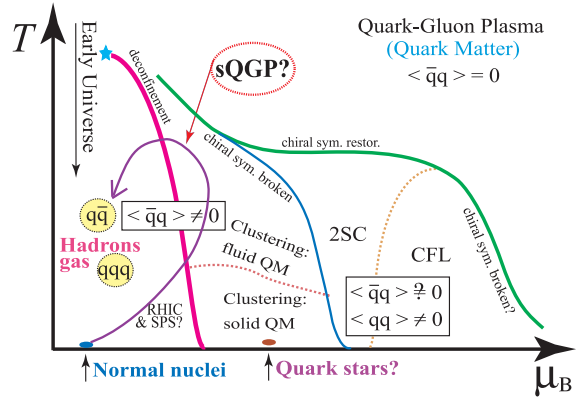}
    \caption{%
Schematic illustration of QCD phase diagram speculated from
different manifestations of astrophysical compact stars. A quark
cluster state is conjectured when chiral symmetry is broken though
quarks are still unconfined. The cold quark matter with quark
clusters should be in a solid state at low temperature.
\label{QCDphase}}
\end{figure}

Interaction can certainly play an essential role to cluster
particles.
For instance, the elements of water are H$_2$O molecules, a cluster
of 10 electrons, 2 protons and 1 oxygen nucleus. However, if the
electromagnetic interaction is turned off, the degenerate electron
chemical potential of ordinary water at a temperature of $\geq
0~^{\rm o}$C is 120 eV, being much higher than 13.6 eV, the
interaction energy between electron and proton in the ground state.
For cold quark matter at $3n_0$ density, the distance between quarks
is $\sim 0.9$ fm $\gg$ the Planck scale $\sim 10^{-20}$ fm, and
quarks and electrons can well be approximated as point-like
particles.
If Q$_\alpha$-like clusters are created in the quark
matter~\cite{xu03}, the distance between clusters are $\sim 2$ fm.
We may also estimate the length scale, $\ell$, of quark clusters by
the uncertainty relation, assuming quarks are as dressed as
speculated in Fig.\ref{QCDphase} (the constituent quark mass is
$m_q\sim 300$ MeV) and move non-relativistically in a cluster.
The quarks typically have a kinematic energy of $\sim \hbar^2/(m_q
\ell^2)$, and are bound by color interaction, with energy of $\sim
\alpha_s\hbar c/\ell$. We have then $\ell\sim \hbar c/(\alpha_s m_q
c^2)\simeq 1$ fm if $\alpha_s\sim 1$, and quark clusters could be
considered as classical particles in cold quark matter and would be
in lattices at a lower temperature.
Therefore, we would like to learn QCD dynamics at a scale of $\sim
1$ fm from the dense matter in pulsar-like compact stars.

Alternatively, a crystalline color superconducting phase (super
solid) with rigidity is proposed by calculation in an effective
model~\cite{mrs07}, in which the quarks are certainly not condensed
in position space but in momentum space.
It is surely interesting to experimentally or observationally
distinguish between and to search evidence for possible normal and
super solid states although the latter might be more robust than the
former only from a theoretical point of view.
Star quakes could naturally occur in both normal and super solid
quark stars, and the observations of pulsar glitches~\cite{z04} and
SGR giant flares~\cite{xty06} could qualitatively be understood
(when the solid matter breaks) in both those two models.
Nonetheless, there might still be some quantitative differences
between the predictions (e.g. the post-glitch recovery and the
global stellar structure) in those two models, which are very
necessary for further researches. Additionally, because the normal
solid depends on quark clustering while the super solid on gap
modulating, the interaction behaviors between quarks should be
different, that could be tested in sQGP by the LHC and/or FAIR
experiments.
For instance, one may test the phenomenological interaction-models
for sQGP at laboratory through calculating the quark clusters by the
quark molecular dynamics (qMD)~\cite{qMD}.

An astrophysically conjectured QCD phase diagram, with the inclusion
of strong coupling between quarks, is shown in Fig.~\ref{QCDphase}.
In different locations of the diagram, the vacuum has different
features which can be classified into two types: the
perturbative-QCD (pQCD) vacuum and the nonperturbative-QCD (QCD)
vacuum. The coupling is weak in the former, but is strong in the
latter. Quark-antiquark (and gluons) condensations occur in QCD
vacuum (i.e., the expected value of $\langle {\bar q}q \rangle \neq
0$), but not in pQCD vacuum. The chiral symmetry is spontaneously
broken in the case that the vacuum is changed from pQCD to QCD
vacuums, and (bare) quarks become then massive constituent ones
(dressed quarks).
As is addressed, the idea of quark clustering (quark-molecular)
could be tested in sQGP.

\vspace{2mm}%
\noindent%
{\bf 3. To understand observations in solid quark star models}

Various astrophysical observational phenomena could be understood in
terms of (solid) quark star models, including those that are
challenging in conventional neutron star models.
Two kinds of solid quark matter are possible, and it could be
interesting to observationally distinguish between and search
evidence for normal-solid and super-solid states in the future.

{\em Radio pulsars.}
It is generally suggested that radio pulsars of strange quark matter
should have crusts (with mass $\sim 10^{-5}M_\odot$), being similar
to the outer crusts of neutron stars~\cite{afo86}.
This view was criticized by Xu \& Qiao~\cite{xq98}, who proposed
that {\em bare} strange stars (i.e., strange stars without crusts),
being chosen as the interior of radio pulsars, have three
advantages: (1) the spectral features; (2) the binding energy; and
(3) the core collapse process during supernova. This opens thus a
new window to distinguish quark stars from neutron stars via their
magnetosphere and surface radiation according to the striking
differences between the exotic quark surfaces and the normal matter
surfaces.
Clear drifting sub-pulses suggest that vacuum inner gap acceleration
works on polar caps and both positively and negatively charged
particles should be strongly bound, and a bare quark star model is
attractive for explaining the pulse sequences~\cite{xqz99}, and even
the X-ray polar caps~\cite{yue06}, of PSR 0943+10.

The timing behavior may also favor a solid quark star model.
The observed precessions in PSR B1828-11 and PSR B1642-03 challenge
astrophysicists today to re-consider the internal structure of radio
pulsars~\cite{link03}, since the conventional model involves vertex
pinning and even MHD coupling between the crust and the core. More
precession sources besides radio pulsars are also proposed. While
normal glitches are suggested to be the results of the vertex
pinning effect, the recently discovered slow glitches can hardly be
explained. A solid pulsar with rigidity could solve these problems,
and both normal and slow glitches can then be
modeled~\cite{z04,px08}.
It is shown that torque variability increases with  Reynolds numbers
(and thus spin frequency)~\cite{mp2007}, suggesting that al least
the timing noise in fast radio pulsars would be high. This is in
conflict with the fact that the noise in millisecond pulsars is much
lower than that of normal pulsars.

{\em Dead pulsars.}
With regard to the possible ways of identifying quark stars (e.g. by
the mass-radius relations or the maximum spin frequencies), hard
evidence for quark star may be obtained by studying the surface
conditions since the other avenues are subject to poorly known
microscopic physics. Although the bare quark surface could help us
to understand the radio emission~\cite{xqz99}, it should be direct
and intuitive if one can detect real thermal radiation from a quark
surface.
Because the thermal component can hardly be separated from the
strong  magnetospheric emission of active pulsars, dead pulsars with
negligible magnetospheric components should be the ideal targets (a
pulsar becomes dead when the potential drop in the open field line
region is lower than a critical value $\sim 10^{12}$ V as the pulsar
spins down), i.e. dead pulsars are good ones.

Thanks to the advanced X-ray missions, more and more dead
pulsar-like compact objects are discovered, being classified as CCOs
(central compact objects, in supernova remnants) and DTNs (dim
thermal ``neutron'' stars, not associated with supernova remnants).
One would expect that both thermal X-ray emission radiating from
neutron star atmospheres, and atomic spectral lines that formed
there should have been discovered by {\em Chandra} or {\em
XMM-Newton}. However, no clear atomic feature has been found. Such a
thermal featureless spectrum could be a probe for identifying bare
quark stars~\cite{xq98,xu02}.
The absorption features of 1E 1207 were suggested to be cyclotron
lines soon after the discovery~\cite{xwq03}, and 1E 1207 could still
be a bare quark star, even with a low mass~\cite{xu05}, in a
propeller phase.
Recent timing observation~\cite{gh07} of the magnetic field favors
the electron-cyclotron idea for 1E 1207.
Due to a very high plasma frequency of $\sim 20$ MeV, fluid quark
stars with exposed quark surface could be silver-like spheres in the
X-ray band~\cite{afo86}. Nevertheless, what if quark matter is
solid~\cite{xu03}?
In fact, the well-observed Planck-like thermal spectrum of RX J1856
could be better fitted phenomenologically by a metal-like model in
the solid quark star regime~\cite{zxz04}.

{\em Anomalous X-ray Pulsars (AXPs) and Soft Gamma-Ray Repeaters
(SGRs).}
Spin-power was generally thought to be the only free energy for
pulsar-like compact stars until the discovery of accretion-powered
pulsars in X-ray binaries.
However, the X-ray luminosities of AXPs/SGRs are much higher than
their spindown powers, and no binary companions of them has been
discovered.
AXPs/SGRs are then suggested to be in an accretion propeller phase,
however the very difficulty for this viewpoint is to reproduce the
irregular bursts, even super-flares with peak luminosity $\sim 10^7$
times of the Eddington luminosity.
The elastic energy as well as gravitational energy of solid quark
stars could be new kinds of free energy to power
AXPs/SGRs~\cite{Horvath05,xty06,xu07}.
A solid stellar object would inevitably result in starquakes when
strain energy develops to a critical value, and huge gravitational
and elastic energies would then be released, especially during
accretion.
This is called {\em AIQ} (Accretion-Induced star-Quake) mechanism.

{\em Supernova and Gamma-ray Bursts (GRBs).}
The essential difficulty of reproducing two kinds of astronomical
bursts are challenging today's astrophysicists to find realistic
explosive mechanisms. Besides the puzzling center engines of GRBs,
it is still a long-standing problem to simulate supernovae
successfully in the neutrino-driven explosion model. Nevertheless,
it is evident that both kinds of explosions could be related to the
physics of cold matter at supra-nuclear density.
One of the direct and important consequences could be the low
baryon-loading energetic fireballs formed on quark star surfaces,
which might finally result in both supernova and GRBs.
A one-dimensional supernova calculation shows that the
lepton-dominated fireball supported by a bare quark surface do play
a significant role in the explosion dynamics under a photon-driven
scenario~\cite{cyx07}.
Two kinds of central engines for GRBs could be available if pulsars
are actually solid quark stars (i.e., the SNE-type and SGR-type
GRBs), and stochastic quakes after initial GRBs may be responsible
for the X-ray flares of both types~\cite{xl09}.

{\em Others.}
The ultra high energy cosmic rays detected via air-showers could be
actually strangelets~\cite{xw03} since strangelets can behave as
cosmic rays beyond the GZK-cutoff and could be effectively
accelerated in pulsar magnetospheres.
Part of the radio pulsar timing noise could reflect ultra-compact
object binaries~\cite{gong05}, two bare quark stars (or planets) in
close binary systems.
We cannot yet rule out that some precession pulsars are torqued by
quark planets~\cite{lyx07}.
It is also interesting to search for low-mass quark stars (or
planets), especially with masses of $\sim (10^{-1}-10^{-3})M_\odot$,
in white dwarf binaries~\cite{xu05}.

\vspace{2mm}%
\noindent%
{\bf 4. Conclusions}

In this paper it is argued that clustering, rather than color
super-conducting, occurs in cold quark matter at realistic baryon
densities. Cold quark matter is then conjectured to be in a solid
state.
Possible evidence for quark stars are summarized.

\vspace{2mm}%
\noindent%
\small{{\em Acknowledgments}:
We thank Profs. Vivian de la Incera and Efrain Ferrer for their
substantial help to improve the language, and an anonymous referee
for valuable suggestions, and acknowledge the contributions by our
colleagues at the pulsar group of PKU. This work is supported by
NSFC (10573002, 10778611), the National Basic Research Program of
China (Grant 2009CB824800), and by LCWR (LHXZ200602).}

\vspace{2mm}%
\noindent%

\end{document}